\documentclass{IEEEtran}

\usepackage{graphicx}
\usepackage{cite,url}

\usepackage[left=2cm,right=2cm,top=2.5cm,bottom=2.5cm]{geometry}

\begin{document}

\title{Cryptanalysis of an MPEG-video Encryption Scheme Based on Secret Huffman
Tables\thanks{The first author was partially supported by the
Alexander von Humboldt Foundation, Germany, and by The Hong Kong
Polytechnic University's Postdoctoral Fellowships Scheme under grant
no. G-YX63. The work of K.-T. Lo was supported by the Research
Grants Council of the Hong Kong SAR Government under Project no.
523206 (PolyU 5232/06E).}\thanks{Shujun Li is the corresponding
author. Contact him via his personal web site:
\texttt{http://www.hooklee.com}}}

\author{Shujun Li\thanks{Shujun Li is with FernUniversit\"{a}t in Hagen, Lehrgebiet
Informationstechnik, Universit\"{a}tsstra{\ss}e 27, 58084 Hagen,
Germany.}, Guanrong Chen\thanks{Guanrong Chen is with the Department
of Electronic Engineering, City University of Hong Kong, 83 Tat Chee
Avenue, Kowloon Tong, Hong Kong SAR, China.}, Albert
Cheung\thanks{Albert Cheung is with the Department of Building and
Construction and Shenzhen Applied R\&D Centres, City University of
Hong Kong, 83 Tat Chee Avenue, Kowloon Tong, Hong Kong SAR, China.}
and Kwok-Tung Lo\thanks{Kwok-Tung Lo is with the Department of
Electronic and Information Engineering, The Hong Kong Polytechnic
University, Hung Hom, Kowloon, Hong Kong SAR, China.}}

\maketitle

\begin{abstract}
This paper studies the security of a recently-proposed MPEG-video
encryption scheme based on secret Huffman tables. Our cryptanalysis
shows that: 1) the key space of the encryption scheme is not
sufficiently large against divide-and-conquer (DAC) attack and
known-plaintext attack; 2) it is possible to decrypt a cipher-video
with a partially-known key, thus dramatically reducing the
complexity of the DAC brute-force attack in some cases; 3) its
security against the chosen-plaintext attack is very weak. Some
experimental results are included to support the cryptanalytic
results with a brief discuss on how to improve this MPEG-video
encryption scheme.\\

\end{abstract}

\section{Introduction}
\label{section:Introduction}

The extensive use of digital images and videos in today's digital
world makes the security and privacy issues more and more important.
To fulfill such an increasing demand, various encryption algorithms
have been proposed in recent years as possible solutions to content
protection of digital images and videos
\cite{Cheng:PartialImageEncryption:IEEETSP2000,
Qiao:ComparisonMPEGEncryption:CG98, Shi:MPEGEncryption:MMTA2004,
Furht:MultimediaSecurity:Handbook2004,
Furht:MultimediaSecurity:Book2005,
Ahl:ImageVideoEncryption:Book2005, Zeng:MultimediaSecurity:Book2006,
Wu&Kuo:MHTEncryption:IEEETMM2004,
Zeng:VideoScrambling:IEEETCASVT2002,
Zeng:VideoScrambling:IEEETMM2003, Wu:JointMMEncryption:IEEETIP2006},
among which MPEG videos attract special attention due to its
prominent prevalence in consumer electronic markets
\cite{MPEG1-ISOStandard, MPEG2-ISOStandard, MPEG4-ISOStandard}. As
an important way of designing MPEG-video encryption schemes, secret
Huffman tables have been suggested in some designs
\cite{Shi:SecretHuffmanCoding:MM98, Shi:MPEGEncryption:MMTA2004,
Kankanhalli:VideoScrambler:IEEETCE2002,
Wu&Kuo:AudiovisualEncryption:SPIE2001,
Wu&Kuo:EntropyCodecEncryption:SPIE2001,
Xie&Kuo:MHTEncryption:SPIE2003, Wu&Kuo:MHTEncryption:IEEETMM2004}.

The MPEG-video encryption scheme proposed in
\cite{Shi:SecretHuffmanCoding:MM98} (i.e., Algorithm 1 in
\cite{Shi:MPEGEncryption:MMTA2004}) is a light-weight scheme, which
encrypts the plain-video by shuffling VLC (variable-length coding)
entries of same size in each Huffman table. However, because the bit
length of each VLC codeword does not change, the position of each
VLC codeword in the video stream does not change either. Thus, an
attacker can uniquely locate (and thus determine) all VLC codewords
contained in the cipher-video stream, if the plain-video stream or
an independent part (such as a picture or a slice) is known. Once
all distinct VLC codewords are obtained, the whole secret Huffman
table is uniquely reconstructed and the encryption scheme is broken.
That is, this light-weight scheme is not secure against
known/chosen-plaintext attacks. In addition, as pointed out in
\cite{Shi:MPEGEncryption:MMTA2004}, the key space of this encryption
scheme is very limited (especially for Huffman tables with a small
number of VLC entries), so even a brute-force attack may become
feasible.

The MPEG-video encryption scheme proposed in
\cite{Kankanhalli:VideoScrambler:IEEETCE2002} can be considered as
an enhanced version of that in \cite{Shi:SecretHuffmanCoding:MM98}.
In this scheme, five different Huffman tables are shuffled
separately and the shuffling operations are generalized to work on
VLC entries with different sizes, in the hope that the key space can
be enlarged and the security against plaintext attacks can be
improved. Furthermore, as a second guard on the security, random bit
flipping operations are also introduced to further encrypt each
secret Huffman table.

In \cite{Wu&Kuo:AudiovisualEncryption:SPIE2001,
Wu&Kuo:EntropyCodecEncryption:SPIE2001,
Xie&Kuo:MHTEncryption:SPIE2003, Wu&Kuo:MHTEncryption:IEEETMM2004},
multiple Huffman tables (MHT) are introduced, from which one table
is secretly chosen for the encryption of each VLC codeword. A
so-called ``Huffman tree mutation process" is also proposed in
\cite{Wu&Kuo:AudiovisualEncryption:SPIE2001,
Wu&Kuo:EntropyCodecEncryption:SPIE2001,
Wu&Kuo:MHTEncryption:IEEETMM2004} to derive more candidate Huffman
tables from several original tables. 

This paper mainly focuses on security problems with the MPEG-video
encryption scheme proposed in
\cite{Kankanhalli:VideoScrambler:IEEETCE2002}. Our cryptanalysis
shows that this scheme is not sufficiently secure against DAC
(divide-and-conquer) ciphertext-only attack, partial-key attack and
known-plaintext attack, and very weak against chosen-plaintext
attack.

The rest of this paper is organized as follows. In the next section,
a brief introduction to the MPEG-video encryption scheme under study
is given. Then, the main cryptanalysis results are presented in
detail in Sect.~\ref{section:Cryptanalysis}, with some experimental
results. Finally Section~\ref{section:Improving} gives a discussion
on how to improve the security of the MPEG-video encryption scheme,
and the last section concludes this paper.

\section{MPEG-Video Encryption Scheme Under Study}

Given an uncompressed video as the input, an MPEG\footnote{To keep
the description simpler, here we mainly considers MEPG-1/2 encoder.
For special features of MPEG-4 encoder with respect to MPEG-1/2, see
\cite{MPEG4-ISOStandard} for more details.} encoder compresses the
video stream frame by frame in the following steps
\cite{MPEG1-ISOStandard, MPEG2-ISOStandard}:
\begin{enumerate}
\item (optional, for interlaced MPEG-2 videos only) separate a
frame into two field-pictures: a top field and a bottom field;

\item employ differential encoding and motion compensation
techniques to remove most redundancy existing between adjacent
pictures;

\item divide the current picture into a number of slices, each of
which is composed of one or more $16\times 16$ macroblocks;

\item decompose each macroblock into 6, 8 or 12 blocks of size
$8\times 8$: 4 luminance blocks; 1, 2 or 4 Cb chrominance blocks; 1,
2 or 4 Cr chrominance blocks;

\item perform DCT (discrete cosine transform) for each
$8\times 8$ block, and then quantize the 64 DCT coefficients with a
quantiser marix;

\item transform the $8\times 8$ block into a 1-D vector in one of
two possible zigzag scanning orders, and then represent all 64
quantized DCT coefficients (except DC coefficients in intra-blocks)
with a number of RLE (run-length encoding) pairs;

\item encode most DCT coefficients (i.e., RLE pairs) and all
motion vectors using VLC (variable-length-coding) codewords under
the control of some Huffman tables, and represent other DCT
coefficients with fix-length bits strings.
\end{enumerate}
A Huffman table is a two-column tale for realizing the Huffman
coding algorithm \cite{Huffman:HuffmanCoding:PIRE1952,
HuffmanCoding}, which is an entropy encoding algorithm and
transforms an input value into a VLC-codeword with the following
rule: the more frequently the input value occurs, the shorter the
corresponding VLC-codeword should be, and vice versa. In such a way,
one can compress the input data in a lossless form. There are in
total 15 Huffman tables used in MPEG-2 standards
\cite{MPEG2-ISOStandard} (less tables are used in MPEG-1 standard
\cite{MPEG1-ISOStandard}), among which ten tables are used to encode
data elements in various headers and the following five tables are
used to encode visual information -- DCT coefficients in each block
and motion vectors in each macroblock:
\begin{itemize}
\item Table B-10: for encoding motion vectors;

\item Table B-12: for encoding the bit size of the differential
values of DC coefficients in intra luminance blocks;

\item Table B-13: for encoding the bit size of the differential
values of DC coefficients in intra chrominance blocks;

\item Table B-14: for encoding all DCT coefficients of non-intra
blocks and AC coefficients of intra blocks with
\textit{intra\_vlc\_format}=0, where \textit{intra\_vlc\_format} is
a picture-specific flag defined in MPEG-2 standard (which is always
set to be 0 for MPEG-1 videos);

\item Table B-15 (not used in MPEG-1 standard): for encoding all AC
coefficients of intra blocks with \textit{intra\_vlc\_format}=1.
\end{itemize}
As an example of the Huffman tables, Table~\ref{table:B12} shows
Table B-12 used in MPEG-2 standard.

\begin{table}[!htbp]
\caption{Table B-12 in MPEG-2 standard: Variable length codes for
\textit{dct\_dc\_size\_luminance}}
\centering \label{table:B12}
\begin{tabular}{c|c}
\hline Variable length code &
\textit{dct\_dc\_size\_luminance}\\
\hline 100 & 0\\
\hline 00 & 1\\
\hline 01 & 2\\
\hline 101 & 3\\
\hline 110 & 4\\
\hline 1110 & 5\\
\hline 11110 & 6\\
\hline 111110 & 7\\
\hline 1111110 & 8\\
\hline 11111110 & 9\\
\hline 111111110 & 10\\
\hline 111111111 & 11\\
\hline
\end{tabular}
\end{table}

The encryption scheme proposed in
\cite{Kankanhalli:VideoScrambler:IEEETCE2002} is designed by
encrypting the original Huffman tables, i.e., using different
(secret) Huffman tables to replace the original ones. The
aforementioned five Huffman tables, B-10, B-12, B-13, B-14 and B-15,
are chosen to be encrypted separately. These secret Huffman tables
are derived from the original ones by combining the following two
encryption operations.
\begin{itemize}
\item \textit{Shuffling VLC-codewords}: grouping all VLC-codewords into several
subsets according to their bit lengths, and then randomly shuffling
these VLC-codewords within each subset.

\item \textit{Random bit flipping}: randomly flipping the last bit
of each VLC-codeword, and adjusting (if needed) other VLC-codewords
to keep the prefix rule valid.
\end{itemize}
After encrypting all the five Huffman tables with both of the above
two methods, the bit length of some VLC-codewords should be slightly
changed to enhance the security against plaintext attacks (as
discussed in Sect.~\ref{section:Introduction} of this paper) but
should not be changed too much, to avoid a large influence on
compression efficiency.

In \cite{Kankanhalli:VideoScrambler:IEEETCE2002}, the key space was
estimated by enumerating all ``good" encryption methods\footnote{An
encryption method is ``good" if it can produce unintelligible
images. Note that the number of ``good" encryption methods may be
larger if a looser definition of ``unintelligible images" is used.}
of the shuffling operation and the random bit flipping operation, as
shown in Table~\ref{table:keyspace}.
\begin{table}[!htbp]
\caption{The number of good encryption methods of each Huffman table
(Table 3.6 of \cite{Kankanhalli:VideoScrambler:IEEETCE2002}).}
\centering \label{table:keyspace}
\begin{tabular}{c|c}
\hline Huffman table & Number of good
encryption methods\\
\hline B-10 & $3!$\\
\hline B-12 & $7!\times 2^6$\\
\hline B-13 & $6!\times 2^8$\\
\hline B-14 & $6!$\\
\hline B-15 & $16!$\\
\hline
\end{tabular}
\end{table}
Then, the key space size was calculated to be the product of the
five numbers given in the table:
\[
(3!)\times\left(7!\times 2^6\right)\times\left(6!\times
2^8\right)\times(6!)\times(16!)\approx 5.37\times 10^{27}\approx
2^{92}.
\]
As a result of the large key space, it was claimed that the scheme
is sufficiently secure. In the next section, we will point out that
this claim is not grounded.

In \cite{Kankanhalli:VideoScrambler:IEEETCE2002}, an additional
measure is also suggested to further enhance the security against
plaintext attacks -- reshuffling the Huffman tables after a certain
number of frames. In the following, we will mainly consider the
basic scheme without the reshuffling mechanism. The effect of the
reshuffling mechanism will be discussed later in
Sect.~\ref{section:Improving}.

\section{Cryptanalysis}
\label{section:Cryptanalysis}

In this section, we re-study the security of the aforementioned
MPEG-video encryption scheme based on secret Huffman tables, and
point out that it is not so secure as claimed in
\cite{Kankanhalli:VideoScrambler:IEEETCE2002}, especially against
the chosen-plaintext attack. In this section, the terms and
notations in MPEG-2 standard \cite{MPEG2-ISOStandard} will be used,
except those existing in MPEG-1 videos only. The following terms are
used throughout this section to facilitate the description: 1) the
term ``picture" is used instead of ``frame", since the encryption
scheme is independent of the syntax differences between a picture
and a frame; 2) macroblock is abbreviated as ``MB"; 3) a
self-defined term ``MB header" is used to denote the set of all data
elements occurring before the first encoded block in an MB (if none
of the blocks is coded, the MB header is the MB itself).

\subsection{Ciphertext-only attack}

The ciphertext-only attack is the most common attack in practice,
since in general the communication channels are open to the public,
which means that an attacker can observe a number of ciphertexts and
use them to break an encryption scheme
\cite{Schneier:AppliedCryptography96, MOV:CyrptographyHandbook1996}.
There are two different goals in a ciphertext-only attack:
recovering the plaintexts and recovering the secret key. The latter
means that the encryption scheme can be completely broken. This
paper mainly focuses on the recovery of the secret key, i.e., the
secret Huffman tables in the MPEG-video encryption scheme under
study.

The simplest ciphertext-only attack is to exhaustively search all
possible keys to find the unique correct one, which is called the
brute-force attack \cite{Schneier:AppliedCryptography96,
MOV:CyrptographyHandbook1996}. Here, a criterion is needed to verify
each searched key. For the MPEG-video encryption scheme under study,
the occurrence of syntax errors can serve as such a criterion for
detecting wrong keys (i.e., wrongly guessed Huffman tables). When a
wrong Huffman table is used to decrypt a cipher-video, syntax errors
may occur in the decoding procedure due to (but not limited to) the
following reasons.
\begin{itemize}
\item
As mentioned in Sect.~\ref{section:Introduction} of this paper, to
ensure the security against plaintext attacks, there are at least
two different bit lengths in a Huffman table. However, once the bit
length of a data element (i.e., the bit length of the corresponding
VLC-codeword) is wrong, all the following data elements in the
current slice cannot be correctly decoded. For example, if
$dct\_dc\_size$\footnote{Italic symbols mean data elements defined
in the MPEG-2 standard, or in the MPEG-1 standard if it does not
defined in the MPEG-2 standard. For example, here ``$dct\_dc\_size$"
denotes the bit size of a DC differential value, as defined in Sect.
7.2.1 of \cite{MPEG2-ISOStandard}.} is not decoded correctly in an
intra-block, all the following data elements in the current slice
cannot be correctly located and decoded.

\item
For each Huffman table, not all VLC-codewords are valid. For
example, in Table B-10, all VLC-codewords have less than 7 zero bits
and the prefix ``0000 0010" never occurs.

\item
There always exist 23 continuous zero bits between two adjacent MBs.

\item
There may exist some marker bits (must be ``1") in the bit stream:
\begin{itemize}
\item
(not valid for MPEG-1 videos) when $concealment\_motion\_vectors=1$
in an intra-block, there exists a marker bit in the MB header;

\item
(for MPEG-1 videos only) in D-picture, the last bit of each MB must
be a marker bit named $end\_of\_macroblock$.
\end{itemize}

\item
There exist some constraints on the decoded data elements:
\begin{itemize}
\item
each decoded DCT coefficient should not be out of the range
$[-2048,+2047]$ after quantization;

\item
each decoded motion vector should not be out of the range defined in
Table 7-8 of the MPEG-2 standard \cite{MPEG2-ISOStandard}, and must
be within the reference picture after adding the coordinates of the
predicted MB.

\end{itemize}

\item
There must be an EOB VLC-codeword at the end of each block, before
which the total number of decoded DCT coefficients must not be
greater than 64.

\item
The number of MBs within each picture should not be greater than a
maximal value.

\item
Some slice headers may be skipped when the video is decoded with
wrong Huffman tables, which is forbidden for most videos (for
example, the MPEG-1 video and the MPEG-2 video with a restricted
slice structure).
\end{itemize}

By detecting the above syntax errors occurring in the decoding
procedure, one can distinguish most wrong Huffman tables. In
addition, there generally exist a lot of information redundancies in
each decoded block, MB, slice, picture and between adjacent blocks,
MBs, pictures and frames. Therefore, even when no syntax error is
detected, one can still distinguish the wrong Huffman table if a
meaningless picture or a number of slices are decoded or if there
exists excessive noise between many adjacent blocks.

Now, let us see how frequently these syntax errors occur. To make
things simpler, denote the average probability of occurrence of a
syntax error at each syntax element by $p$. Then, for a picture with
$L$ syntax elements, the probability that at least one syntax error
occurs within this picture will be $P(L)=1-(1-p)^L$. For an MPEG-1
video of size $M\times N$, the number of syntax elements is about
$6MN\lambda$, where $\lambda$ is a factor determined by the average
number of syntax elements in each block and ranges roughly from
$1/64$ to 1. As an example, taking $M=176$, $N=144$, $\lambda=1/64$
and $p=0.001$, one can easily calculate and obtain $L=2376$ and
$P(L)\approx 0.9072$. Since the value of $\lambda$ is generally
larger than $1/64$ and the value of $p$ might not be so small, it is
generally a high-probability event to observe at least one syntax
error within one MPEG-picture. Note that in reality the values of
$p$ and $\lambda$ vary in a wide range due to the following reasons:
1) there are a lot of possibilities of encoding a video within the
framework of MPEG standards; 2) there are some optional data
elements in MEPG standards; 3) the occurrence of some kinds of
syntax elements depends on the contents of the encoded blocks or
other syntax elements; 4) different blocks correspond to different
distributions of DCT coefficients, which directly influence the
values of $\lambda$ and $p$; 5) different Huffman tables (especially
these short VLC-entries) correspond to different value of $p$.
Therefore, instead of estimating the values of $p$ and $L$, the
efficiency of the syntax-error detection process will be shown by
carrying out experiments on some sample videos.

Because the five Huffman tables are used for different parts of the
whole video bit-stream, it is possible to separately guess them one
by one. This means that one can use the so-called divide-and-conquer
(DAC) attack \cite{Schneier:AppliedCryptography96,
MOV:CyrptographyHandbook1996} to break the MPEG-video encryption
scheme. In other words, the key space of the encryption scheme will
be the \textbf{\em sum}, not the \textbf{\em product}, of the
sub-spaces of the five tables. Next, let us see how to separately
break the five Huffman tables by detecting syntax errors in the
video decoding procedure.

\subsubsection{Reconstructing Table B-10}

Following the MPEG-2 standard, Tables B-12/13/14/15 are all
independent of the decoding of the first MB header in a
slice\footnote{All other MBs cannot be located without knowing
Tables B-12/13/14/15.}, which makes the separate reconstruction of
Table B-10 possible. When a wrong Table B-10 is used, the following
syntax errors may occur when the first MB header of a slice is
decoded.
\begin{itemize}
\item
Some decoded motion vectors may be invalid, especially for those MBs
near the picture edge.

\item When $concealment\_motion\_vectors=1$ in an intra-block, the
marker bit in the MB will be wrong (i.e., equal to 0) with a
probability of 0.5 (under the assumption that each bit in the video
stream is distributed uniformly).

\item
When $macroblock\_pattern=1$, ``0000 0000 0" never occurs in
$coded\_block\_pattern$ encoded by Table B-9.
\end{itemize}
Note that in each slice only the first MB header can be used to
detect syntax errors about Table B-10, so the average number of
involved syntax elements in each block (i.e., the value of $\lambda$
corresponding to this table) may be too small, especially for
pictures with a small number of slices and/or slow motion. When such
an event happens, one has to exhaustively search through Table B-10
and other Huffman tables. Experiments showed that this event really
happened for many MPEG videos.

\subsubsection{Reconstructing Table B-14}

Since all DCT coefficients in a non-intra MB are encoded with Table
B-14, syntax errors may occur when continuous non-intra MBs in a
slice are decoded with a wrong Table B-14. Because most MBs in a
P/B-picture are non-intra MBs, the occurrence probability of syntax
errors (i.e., the value of $p$ corresponding to this table) will be
relatively high.

To test how frequently syntax errors of this kind occur in real
attacks, we observed the decoding process by exchanging the
following two VLC-entries in Table B-14 -- ``11s" and ``011s", which
represent RLE-codewords (0,1) and (1,1), respectively\footnote{The
two VLC-codewords were chosen due to their frequent occurrence in
the MPEG-1/2 video stream.}. For a large number of test MPEG-1/2
videos, syntax errors started to occur after a few number of MBs
were decoded. See Fig.~\ref{figure:BreakingB14} for the decoding
results of an MPEG-1 video ``Carphone" (of size $176\times 144$) and
an MPEG-2 video ``Tennis"\footnote{The 1st picture of ``Tennis" is
encoded with $intra\_vlc\_format=1$, while the 2nd picture with
$intra\_vlc\_format=0$.} (of size $704\times 576$). If the whole
Huffman table is heavily shuffled, syntax errors will occur even
more frequently.

\newlength\sfigwidth
\setlength\sfigwidth{0.24\textwidth}

\begin{figure}[!htbp]
\centering
\begin{minipage}{\sfigwidth}
\centering
\includegraphics[width=\textwidth]{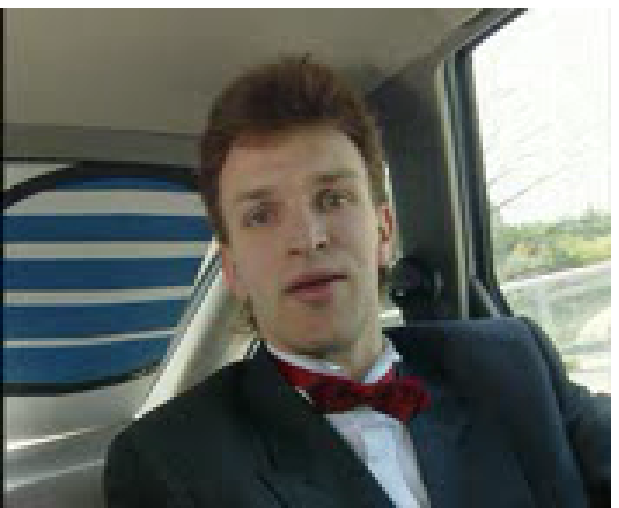}
a)
\end{minipage}
\begin{minipage}{\sfigwidth}
\centering
\includegraphics[width=\textwidth]{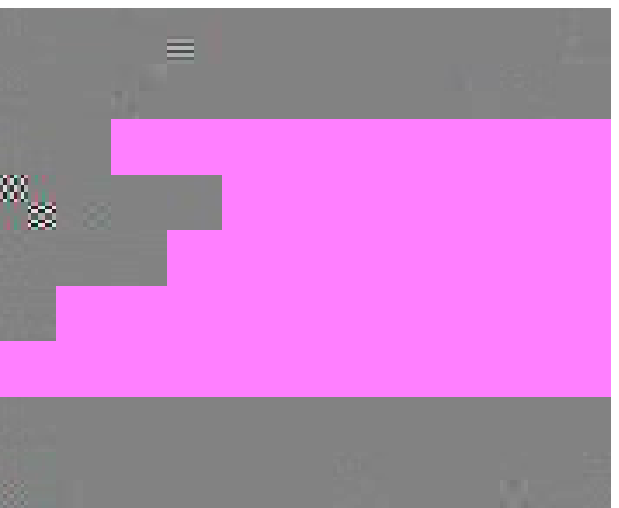}
b)
\end{minipage}
\begin{minipage}{\sfigwidth}
\centering
\includegraphics[width=\textwidth]{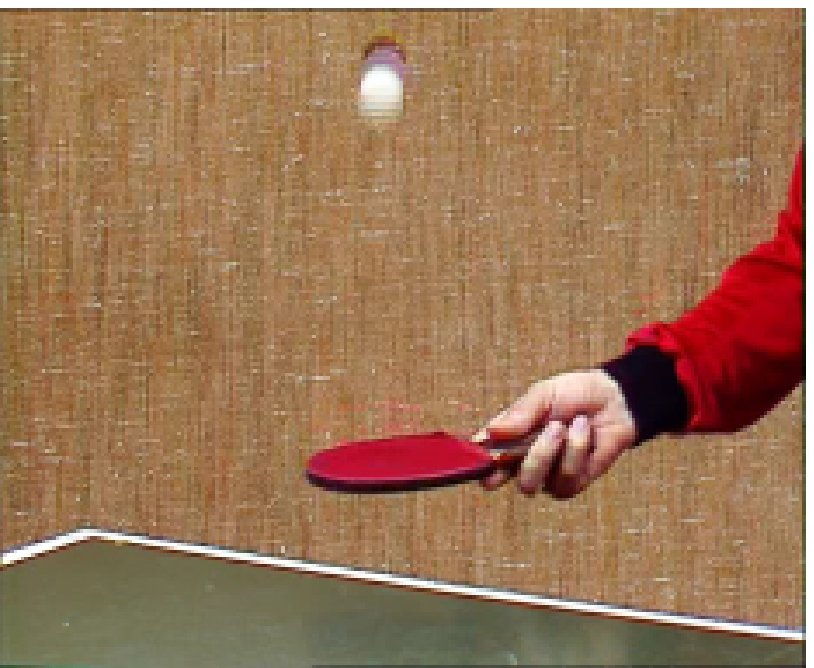}
c)
\end{minipage}
\begin{minipage}{\sfigwidth}
\centering
\includegraphics[width=\textwidth]{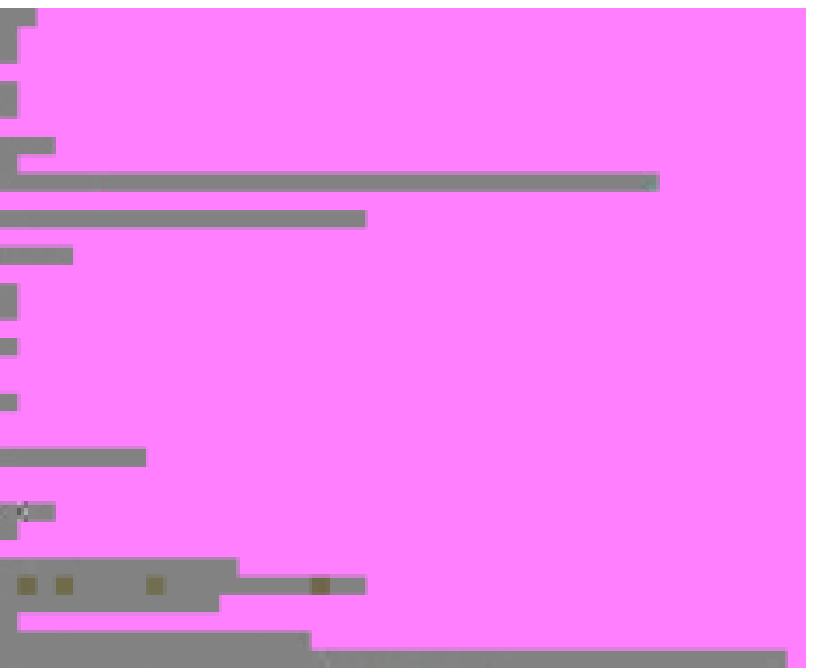}
d)
\end{minipage}
\caption{The decoded results of an MPEG-1 video ``Carphone" and an
MPEG-2 video ``Tennis", when only two VLC-codewords were exchanged
in Table B-14: a) the 2nd picture of ``Carphone"; b) the decoded 2nd
picture of ``Carphone"; c) the 2nd picture of ``Tennis"; d) the
decoded 2nd picture of ``Tennis". The light grey areas in the
decoded frames denote decoding failures caused by syntax errors (the
same hereinafter). Note that the decoded pictures are displayed as
raw data (i.e., the differential pictures) since the reference
I-pictures are normally unknown at this stage.}
\label{figure:BreakingB14}
\end{figure}

\subsubsection{Reconstructing Table B-12}

Once Table B-14 is reconstructed, Table B-12 can be further
exhaustively searched in intra MBs with $intra\_vlc\_format=0$. If
the attacker can get at least one MPEG-1 cipher-video, Table B-12
can always be broken separately (Table B-15 is not used in MPEG-1
videos, which means $intra\_vlc\_format=0$). If all intra MBs in all
known plain-videos are encoded with $intra\_vlc\_format=1$, Table
B-12 has to be exhaustively searched together with Table B-15 (see
below), which is generally a rare event when an attacker can collect
a number of cipher-pictures to carry the ciphertext-only attack.

\begin{figure}[!htbp]
\centering
\begin{minipage}{\sfigwidth}
\centering
\includegraphics[width=\textwidth]{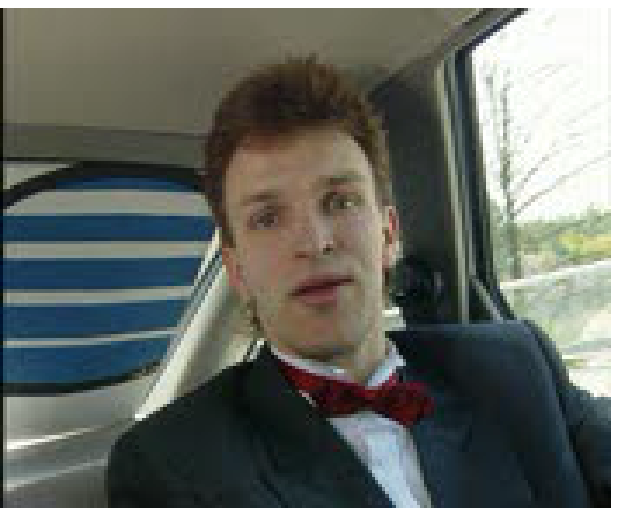}
a)
\end{minipage}
\begin{minipage}{\sfigwidth}
\centering
\includegraphics[width=\textwidth]{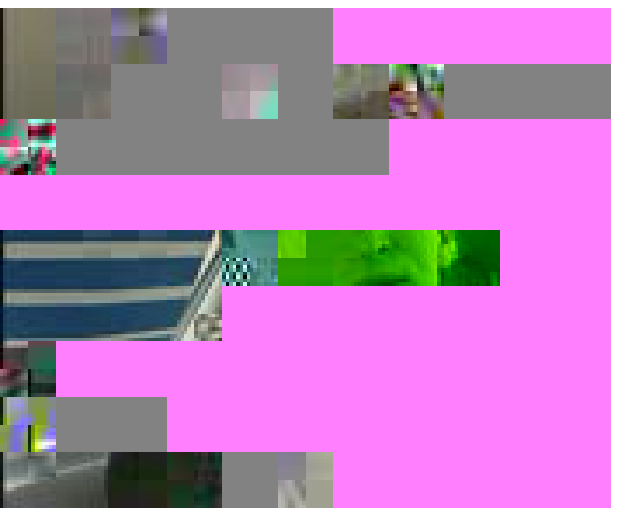}
b)
\end{minipage}
\begin{minipage}{\sfigwidth}
\centering
\includegraphics[width=\textwidth]{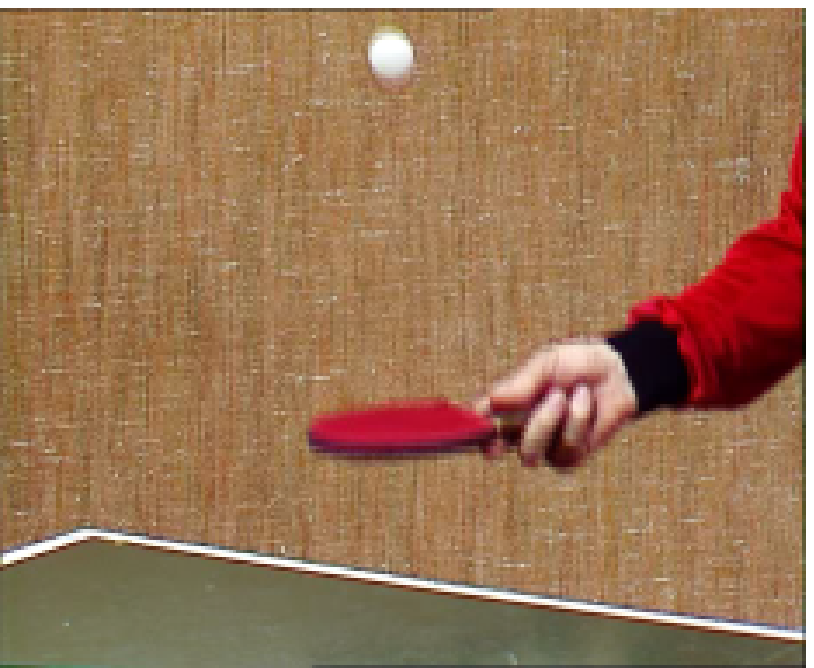}
c)
\end{minipage}
\begin{minipage}{\sfigwidth}
\centering
\includegraphics[width=\textwidth]{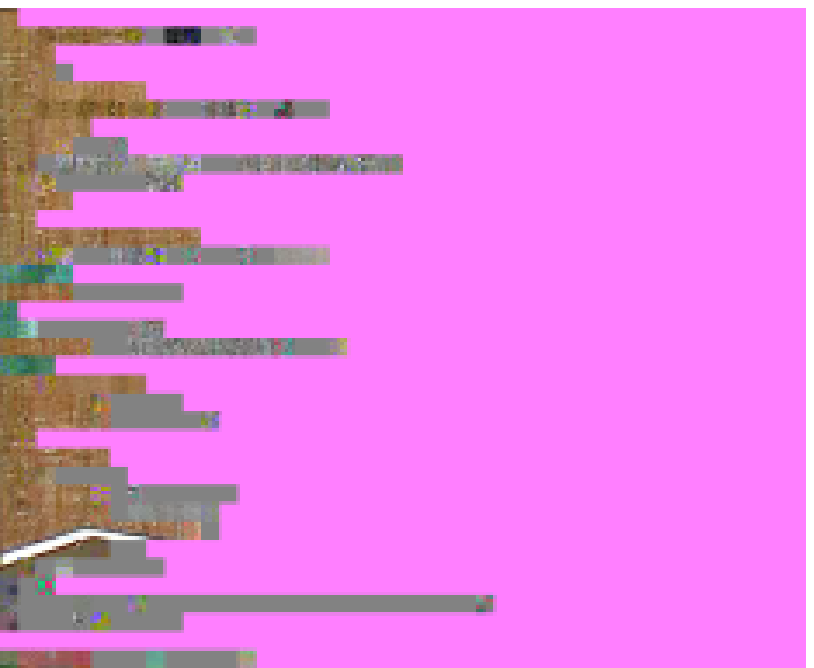}
d)
\end{minipage}
\caption{The decoded results of the MPEG-1 video ``Carphone" and the
MPEG-2 video ``Tennis", when only two VLC-codewords were exchanged
in Table B-12: a) the 1st picture of ``Carphone"; b) the decoded 1st
picture of ``Carphone"; c) the 1st picture of ``Tennis"; d) the
decoded 1st picture of ``Tennis".} \label{figure:BreakingB12}
\end{figure}

In the case that only two VLC-codewords, ``00" and ``01", in Table
B-12 were swapped, we tested the decoding results of some MPEG-1/2
videos. Figure~\ref{figure:BreakingB12} gives the results of the
MPEG-1 video ``Carphone" and the MPEG-2 video ``Tennis". One can see
that the syntax errors still occur very frequently. Note that the
swapped VLC-codewords have the same bit length, so a stronger
shuffling of Table B-12 shall cause much more syntax errors.

\subsubsection{Reconstructing Table B-15}

If Table B-12 has been successfully guessed, Table B-15 can be
exhaustively searched in luminance blocks of intra MBs with
$intra\_vlc\_format=1$, just like the case of reconstructing Table
B-14. If Table B-12 cannot be separately broken, Tables B-12 and
B-15 have to be exhaustively searched together.

\begin{figure}[!htbp]
\centering
\begin{minipage}{\sfigwidth}
\centering
\includegraphics[width=\textwidth]{Tennis1}
a)
\end{minipage}
\begin{minipage}{\sfigwidth}
\centering
\includegraphics[width=\textwidth]{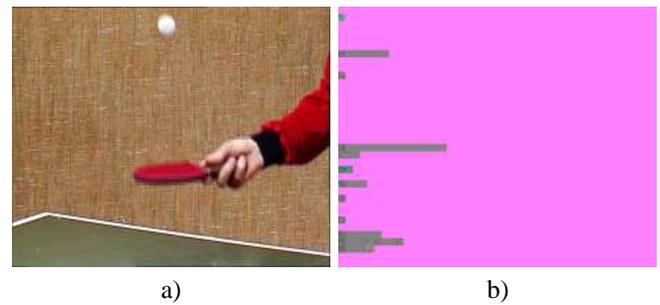}
b)
\end{minipage}
\caption{The decoded result of the MPEG-2 video ``Tennis", when only
two VLC-codewords were exchanged in Table B-15: a) the 1st picture
of the original video; b) the decoded 1st picture.}
\label{figure:BreakingB15}
\end{figure}

By swapping ``10s" and ``010s", which represent (0,1) and (1,1),
respectively, in Table B-15, we tested the decoding results of some
MPEG-2 videos (note that this table is not used in the MPEG-1
standard). Figure~\ref{figure:BreakingB15} gives the results of the
MPEG-2 video ``Tennis". It can be seen again that many syntax errors
still occur.

\subsubsection{Reconstructing Table B-13}

After Tables B-12, 14 and 15 are broken, Table B-13 can be
exhaustively searched in chrominance blocks of intra MBs. If there
are intra MBs with $intra\_vlc\_format=0$, Table B-13 can be
exhaustively broken immediately after Table B-14 is broken, without
knowing Table B-15.

\begin{figure}[!htbp]
\centering
\begin{minipage}{\sfigwidth}
\centering
\includegraphics[width=\textwidth]{Carphone1}
a)
\end{minipage}
\begin{minipage}{\sfigwidth}
\centering
\includegraphics[width=\textwidth]{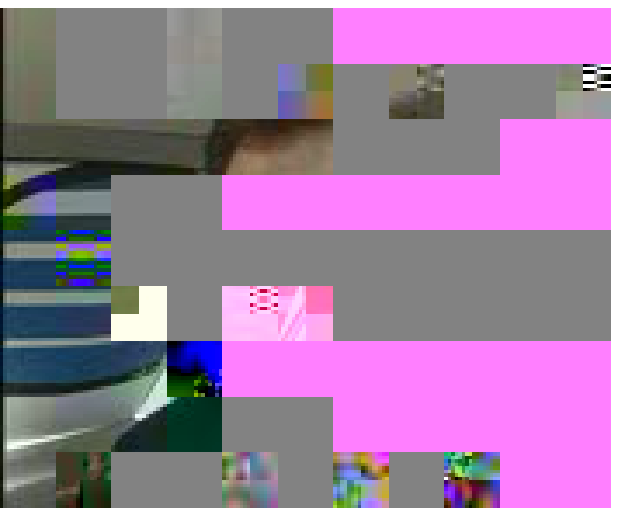}
b)
\end{minipage}
\begin{minipage}{\sfigwidth}
\centering
\includegraphics[width=\textwidth]{Tennis1}
c)
\end{minipage}
\begin{minipage}{\sfigwidth}
\centering
\includegraphics[width=\textwidth]{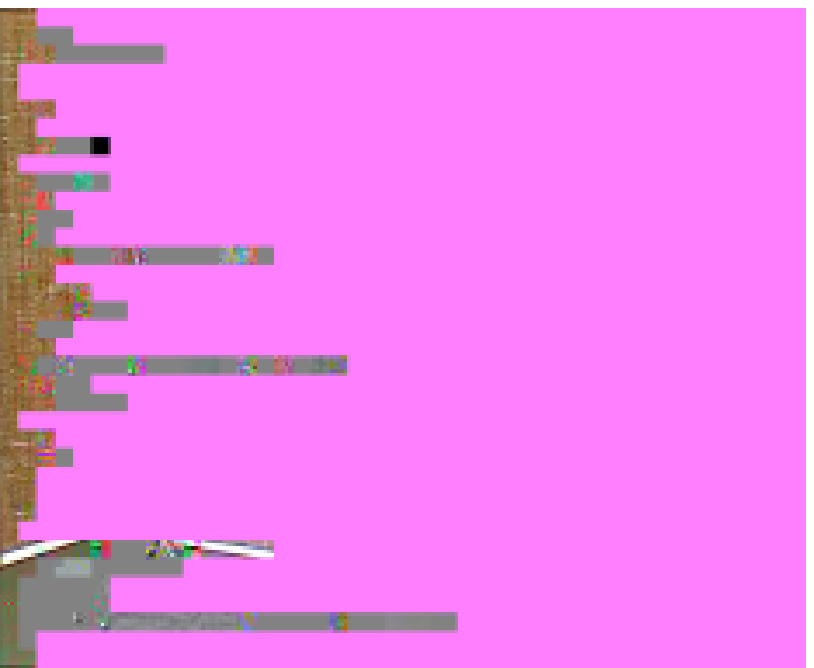}
d)
\end{minipage}
\caption{The decoded results of the MPEG-1 video ``Carphone" and the
MPEG-2 video ``Tennis", when only two VLC-codewords were exchanged
in Table B-13: a) the 1st picture of ``Carphone"; b) the decoded 1st
picture of ``Carphone"; c) the 1st picture of ``Tennis"; d) the
decoded 1st picture of ``Tennis".} \label{figure:BreakingB13}
\end{figure}

By exchanging two VLC-codewords, ``01" and ``10", in Table B-13, we
tested the decoding results of some MPEG-1/2 videos. The results
corresponding to the MPEG-1 video ``Carphone" and the MPEG-2 video
``Tennis" are shown in Fig.~\ref{figure:BreakingB13}. Again, many
syntax errors can be observed. Due to a similar reason in the case
of Table B-12, even more syntax errors are expected in a real
shuffling of Table B-13.

Finally, based on the above analysis and experimental results, let
us estimate the complexity of the above DAC attack under four
different conditions as follows:
\begin{itemize}
\item \textit{when Table B-10 is separately reconstructed}:
\begin{itemize}
\item \textit{when Table B-12 is separately searched}:
$(3!)+(7!\times 2^6)+(6!\times 2^8)+(6!)+(16!)\approx 2^{44.3}$;

\item \textit{when Table B-12 is searched together with Table
B-15}: $(3!)+(7!\times 2^6)\times(16!)+(6!\times 2^8)+(6!)\approx
2^{62.5}$;
\end{itemize}

\item \textit{when Table B-10 is not separately reconstructed}:
\begin{itemize}
\item \textit{when Table B-12 is separately searched}:
$(3!)\times((7!\times 2^6)+(6!\times 2^8)+(6!)+(16!))\approx
2^{46.8}$;

\item \textit{when Table B-12 is searched together with Table
B-15}: $(3!)\times((7!\times 2^6)\times(16!)+(6!\times
2^8)+(6!))\approx 2^{65.1}$.
\end{itemize}
\end{itemize}
One can see that in all cases the complexity will be much smaller
than the one given in \cite{Kankanhalli:VideoScrambler:IEEETCE2002}:
$(3!)\times(7!\times 2^6)\times(6!\times
2^8)\times(6!)\times(16!)\approx 2^{92}$. Considering that Table
B-12 can be separately searched in most cases, this Huffman table
based encryption can be easily broken with a practically small
complexity.

As discussed above, since each picture of a cipher-video contains a
large number of VLC data elements, syntax errors about Tables B-12,
B-13, B-14 and B-15 will generally occur in each picture with a
relatively high probability. That is, except Table B-10, all other
Huffman tables can be separately broken in most cases, where note
that Table B-10 makes the smallest contribution to the whole key
space among all the five tables. So, it is expected that only a few
number of cipher-pictures (or even a few number of consecutive
cipher-slices) is enough to carry out the above DAC attack
effectively.

\subsection{Partial-key attack}

There exists another serious defect in the MPEG-video encryption
scheme under study: with a partially-recovered key (i.e., partial
entries of the Huffman tables), it is still possible to decrypt a
cipher-video, if this cipher-video does not contain any undetermined
entries in the Huffman tables. For example, if the ``Escape" entry
in Table B-14 or B-15 is not reconstructed, one can still use the
partially-reconstructed Huffman tables to decrypt a cipher-video
that does not contain Escape DCT coefficients. Thus, the complexity
of finding a practicable Huffman table is generally smaller than the
complexity of exactly recovering the whole Huffman table itself,
where the word ``practicable" means that a partially-reconstructed
Huffman table can be used to decrypt some specific target
ciphertexts, though not all possible ciphertexts.

Furthermore, Table B-15 is a very special Huffman table, since it
never occurs in MPEG-1 videos and may not occur in some MPEG-2
videos as well. So, Table B-15 can be simply neglected when breaking
such videos, which means that Table B-12 can always be searched
separately. In this case, the key space under the DAC
ciphertext-only attack becomes even smaller:
\begin{itemize}
\item \textit{when Table B-10 is separately reconstructed}:
$(3!)+(7!\times 2^6)+(6!\times 2^8)+(6!)=507606\approx 2^{19.0}$;

\item \textit{when Table B-10 is not separately reconstructed}:
$(3!)\times((7!\times 2^6)+(6!\times 2^8)+(6!))=3045600\approx
2^{21.5}$.
\end{itemize}
One can see that the key space is so small that it is even possible
to find the secret Huffman tables within seconds on a PC with a 1GHz
CPU (note that 1G$=2^{30}\gg 2^{21.5}$).

\subsection{Chosen-plaintext attack}

The chosen-plaintext attack is a very strong attack, in which one
can (intentionally) choose some plaintexts and observe the
corresponding ciphertexts to break an encryption scheme
\cite{Schneier:AppliedCryptography96, MOV:CyrptographyHandbook1996}.
With the help of some chosen plaintexts and ciphertexts, it is
possible to directly determine the secret Huffman tables without
exhaustively guessing them in all possible candidates. In the
following, we show how to choose the data elements in a plain-video
to carry out a successful chosen-plaintext attack.

\subsubsection{Reconstructing Table B-10}

Choose a P/B-picture so that all MBs are encoded as ``Not Coded",
i.e., only MB headers occur in this picture and all blocks are
skipped. Given any two MBs in a slice, one can easily locate the 23
zero bits between the two MBs and all data elements occurring before
motion vectors in the first MB header. Note that the data elements
in the second MB header can be intentionally chosen to facilitate
such a locating procedure. Then, one can extract a bit segment from
the first MB header, which is composed of two or four motion
vectors. In the extracted bit segment, the values of
$motion\_residuals$, dmvectors and the sign bits of the motion
vectors can be chosen to uniquely distinguish each $motion\_code$,
i.e., each VLC-codeword encoded with the secret Table B-10. If
necessary, $f\_code[r][s]$ can also be intentionally chosen to help
the extraction of the VLC-encoded $motion\_codes$. By choosing the
values of 17 consecutive $motion\_codes$ to be 0, ..., 16,
respectively, all entries in Table B-10 can be uniquely determined.
Apparently, to completely break Table B-10, only a few MBs (not a
full picture) are needed.

\subsubsection{Reconstructing Table B-14}

After reconstructing Table B-10, one can continue to break Table
B-14 in non-intra MBs. The entries in Table B-14 can be
reconstructed as follows:
\begin{itemize}
\item
\textit{EOB and Escape entries}: At the beginning of the first MB in
a slice, choose the following two plain-blocks (in the zigzag
scanning order, the same hereinafter): block \#1 --
``$\overbrace{0,\cdots,0}^{run_e}$, $level_e$, 0, $\cdots$", block
\#2 -- ``$\overbrace{0,\cdots,0}^{run_e}$, $-level_e$, 0, $\cdots$",
where $(run_e,level_e)$ forms an Escape RLE-codeword (i.e., not a
valid entry in Table B-14). Then, the video bitstream corresponding
to the two plain-blocks will be: ``Escape, $run_e$, $level_e$, EOB,
Escape, $run_e$, $-level_e$, EOB". By choosing $run_e$, $level_e$ to
be some special values and locating the two fixed-length bit
strings: ``$run_e$, $level_e$" and ``$run_e$, $-level_e$", it is
easy to determine the Escape and EOB VLC-codewords of Table B-14.

\item
\textit{Other entries}: Choose a block with two RLE-codewords:
``$\overbrace{0,\cdots,0}^{run}$, $level$,
$\overbrace{0,\cdots,0}^{run_e}$, $level_e$, 0, $\cdots$" where
$(run,level)$ is a valid entry in Table B-14 and $(run_e,level_e)$
is an Escape RLE-codeword. By choosing $(run_e,level_e)$ properly,
one can easily extract the VLC-codeword of $(run,level)$ from the
bit stream. Repeating this process for all RLE-codewords in Table
B-14, one can completely recover the whole Table B-14.
\end{itemize}
In the above reconstruction process, the number of required blocks
is equal to the number of entries in Table B-14, which means only
tens of MBs.

\subsubsection{Reconstructing Tables B-12 and B-13}

Since AC coefficients of intra-blocks are encoded in a similar way
to the motion vectors, the method of reconstructing Table B-10 can
also be used to break Tables B-12 and B-13. To break the entry
corresponding to $dct\_dc\_size=s$, choose an intra-block as
follows: ``$level$, $\overbrace{0,\cdots,0}^{63}$", where $level$
has $s$ significant bits. Then, the video bitstream corresponding to
this block will be ``$dct\_dc\_size$, $dc\_dct\_differential$, EOB".
Since EOB and $dc\_dct\_differential$ are both known, it is easy to
determine the VLC-encoded $dct\_dc\_size$. Given 12 luminance blocks
with $s=0\sim 11$, Table B-12 can be completely reconstructed.
Similarly, given 12 chrominance blocks with $s=0\sim 11$, Table B-13
can be completely reconstructed. Apparently, the number of required
chosen MBs is 3, 6 or 12 according to the value of $chroma\_format$
(4, 2 or 1).

\subsubsection{Reconstructing Table B-15}

After reconstructing Tables B-12 and B-13, one can break Table B-15
by choosing some intra-blocks, in the same way of reconstructing
Table B-14.

As a whole, one can see that only tens of chosen MBs are enough to
break all the five secret Huffman tables. When the picture is not
too small, this means that only one chosen picture is enough to
break the whole encryption scheme. So the MPEG-video encryption
scheme is very weak against chosen-plaintext attack.

\subsection{Known-plaintext attack}

The known-plaintext attack is a weak version of chosen-plaintext
attack, in which one can only passively observe a number of
plaintexts and the corresponding ciphertexts to break an encryption
scheme \cite{Schneier:AppliedCryptography96,
MOV:CyrptographyHandbook1996}.

Apparently, if some chosen blocks mentioned in the above-mentioned
chosen-plaintext attack are observed in a known-plaintext attack,
the corresponding entries in the involved Huffman tables can be
immediately reconstructed. In addition, the FLC (fixed-length
coding) data element following $dct\_dc\_size$ can be used in
known-plaintext attack to uniquely locate $dct\_dc\_size$. This
means that Tables B-12 and B-13 may be reconstructed directly. Also,
the reconstructed VLC-entries in a Huffman table can be used to
locate other undetermined VLC-entries and to detect wrong candidate
entries, which can further reduce the attack complexity. Generally
speaking, the complexity of the known-plaintext attack shall be much
smaller than the complexity of the DAC brute-force attack, though
more plain-MBs are required as compared with chosen-plaintext
attack.

\section{Improving the MPEG-Video Encryption Scheme}
\label{section:Improving}

Though the main focus of this cryptanalysis paper is to point out
some security flaws of the MPEG video encryption scheme proposed in
\cite{Kankanhalli:VideoScrambler:IEEETCE2002}, in this section we
give a brief discussion on how to improve the security of the MPEG
video encryption scheme under study, hoping that more sequential
studies in this research area can be motivated.

To improve the security of the MPEG-video encryption scheme, a
simple way is to change the Huffman tables frequently. In
\cite{Kankanhalli:VideoScrambler:IEEETCE2002}, it was suggested to
reshuffle the Huffman tables after certain number of frames.
Generally speaking, these reshuffling operations might be enough to
provide an acceptable resistance against ciphertext-only attack.
However, even reshuffling these Huffman tables frame by frame is
generally not sufficient for the security against the above
chosen-plaintext attack, since a few number of slices may be enough
to break the secret Huffman tables. From the most conservative point
of view, one has to reshuffle the Huffman tables for each
VLC-codeword. Such a heavy reshuffling process will dramatically
reduce the speed of the whole system and become impractical in many
real applications.

Another possible solution is to use multiple Huffman tables as
suggested in \cite{Wu&Kuo:AudiovisualEncryption:SPIE2001,
Wu&Kuo:EntropyCodecEncryption:SPIE2001,
Xie&Kuo:MHTEncryption:SPIE2003, Wu&Kuo:MHTEncryption:IEEETMM2004}.
As a typical implementation of this kind of MHT-encryption schemes,
a stream cipher (or a secure PRNG) is adopted to determine the
secret Huffman table from multiple candidate tables for each
VLC-codeword. However, as is well known in cryptology, a stream
cipher is not secure against plaintext attacks if the key is reused
to encrypt more than two plain messages. Thus, in real applications,
to further guarantee the security of this MHT-encryption scheme
against plaintext attacks, one of the following practical measures
may be adopted to avoid potential security defects that may arise
from the embedded stream cipher.

\begin{itemize}
\item
\textit{Avoiding reuse of the same secret key to encrypt two videos,
i.e., changing the secret key for different videos}: this measure
has to be used together with a key management system and may not be
very useful in low-cost video applications (such as storage of
private videos in personal computers and mobile devices).

\item
\textit{Assigning a unique ID (UID) for each video (by the
manufacturer or by the end user), and then using the UID to
initialize the stream cipher together with the secret key}: this
measure can be considered as a special case of the above measure,
but it can work without a key management system.

\item
\textit{Using plaintext or ciphertext feedback to make the stream
cipher dependent on the whole plain-video}: this measure will bring
up an error-propagation problem, but can work well in error-free
environments.

\item
\textit{Combining the secret selection of the Huffman tables with a
block cipher to construct a product cipher}: the block cipher should
be sufficiently simple and fast, and can even be insecure when used
separately (for example, with a small block size). This measure may
lead to some new designs of MPEG video encryption schemes and needs
future studies.
\end{itemize}

Finally, note that the security defects about the DAC brute-force
attack and the partial-key attack cannot be essentially avoided even
with the above countermeasures, since they are actually caused by
the inherent feature of an MPEG-video's syntax structure. This
implies that only using secret Huffman tables is not sufficient to
provide an acceptable security level for all MPEG-videos. Some more
powerful techniques, such as secret permutation of DCT coefficients
and encryption of VLC indices
\cite{Zeng:VideoScrambling:IEEETCASVT2002,
Zeng:VideoScrambling:IEEETMM2003, Wu:JointMMEncryption:IEEETIP2006},
have to be introduced to achieve such a goal for MPEG-video
encryption. In future, we will investigate how to combine different
encryption methods to design MPEG-video encryption scheme with high
level of security.

\section{Conclusions}

This paper has analyzed the security of a recently-proposed
MPEG-video encryption scheme, which bases its security of the use of
some secret Huffman tables. As a result, it is found that the scheme
is not sufficiently secure against DAC (divide-and-conquer)
brute-force attack and known-plaintext attack, and is very weak
against the chosen-plaintext attack. Another serious security defect
of this scheme is that a partially-known key may be used to decrypt
some cipher-videos, which further causes a reduction of the key
space. Based on our cryptanalytic results, a brief discussion is
also given on how to further improve the security of the MPEG-video
encryption scheme under study.

\bibliographystyle{IEEEtran}
\bibliography{Huffman}

\end{document}